\documentclass[9pt,twocolumn,twoside]{article}

\usepackage{amsmath}
\usepackage{amssymb}
\usepackage{graphicx}
\usepackage{stfloats}
\usepackage{placeins}
\usepackage{fullpage}
\usepackage{fixltx2e}
\usepackage[superscript,biblabel]{cite}
\usepackage{setspace}
\usepackage{caption}
\usepackage{abstract}
\usepackage{authblk}
\usepackage[margin=0.6in]{geometry}
\title{\Huge \textbf{Focusing light by wavefront shaping through disorder and nonlinearity}}

\author[1]{Hadas Frostig}
\author[1]{Eran Small}
\author[1]{Anat Daniel}
\author[1]{Patric Oulevey}
\author[1,2]{Stanislav Derevyanko}
\author[1,*]{Yaron Silberberg}

\affil[1]{\footnotesize Department of Physics of Complex Systems, Weizmann Institute of Science, Rehovot 7610001, Israel.}
\affil[2]{Department of Electrical and Computer Engineering, Weizmann Institute of Science, Ben-Gurion University of the Negev, Beer Sheva 8499000, Israel}
\affil[*]{Corresponding author: yaron.silberberg@weizmann.ac.il}

\begin{document}
\twocolumn[
  \maketitle
\begin{onecolabstract}
\normalsize \textbf{Wavefront shaping is a powerful technique that can be used to focus light through scattering media, which can be important for imaging through scattering samples such as tissue. The method is based on the assumption that the field at the output of the medium is a linear superposition of the modes traveling through different paths in the medium. However, when the scattering medium also exhibits nonlinearity, as may occur in multiphoton microscopy, this assumption is violated and the applicability of wavefront shaping becomes unclear. Here we show, using a simple model system with a scattering layer followed by a nonlinear layer, that with adaptive optimization of the wavefront light can still be controlled and focused through a scattering medium in the presence of nonlinearity. Notably, we find that moderate positive nonlinearity can serve to significantly increase the focused fraction of power, whereas negative nonlinearity reduces it.}
\vspace{10mm}
\end{onecolabstract}]

\section{Introduction}
A major limitation of optical imaging is the inability to image deep into inhomogeneous media. Inhomogeneity causes scattering, which randomizes the direction of propagation of the light and prevents focusing or imaging with a lens. Recently, Vellekoop and Mosk have demonstrated that optimization of the incident wavefront using a spatial light modulator (SLM) can be used to partially restore the diffraction-limited focus \cite{Vellekoop2007a}. Conceptually, each pixel of the SLM can be viewed as the source of a different speckle pattern, and by devising a clever phase mask all these patterns can be made to constructively interfere at a given point at the output plane, enhancing its intensity by a factor that is approximately equal to the number of pixels. This successful demonstration prompted a considerable amount of research on the feasibility and applicability of wavefront shaping in many fields \cite{Cizmar2010,Vellekoop2010,Popoff2010,vanPutten2011,Xu2011,Katz2011a,Aulbach2011,McCabe2011,Choi2011}, including imaging through scattering tissue \cite{Yaqoob2008,Wang2012,Katz2012}. One challenge in the application of this technique to imaging through scattering media is that while the brightness of the focus can be increased by 2-3 orders of magnitude, the power contained in the obtained focus is only a small fraction of the total available power. This limitation is imposed by the finite number of degrees of control that can be utilized in practice in wavefront shaping, which is typically significantly smaller than the number of transmission modes through the scattering medium, even when it is thin \cite{Mosk2012}. The rest of the light remains in the form of a speckle pattern, which creates a strong background signal and may completely overwhelm the signal from the obtained focus. Additionally, since the power at the focus is limited, the total intensity of the incident light must be significantly increased to compensate and may damage the sample.

When the medium exhibits nonlinearity in addition to inhomogeneity, the field at the output plane can no longer be described as a linear superposition of speckle fields originating from the SLM pixels, since these fields interact as they propagate \cite{Barsi2009,Goy2011}. As a result, commonly-used methods of focusing through scattering media may become ineffective. Transmission matrix based methods, for example, rely on the linear superposition principle and therefore may be inadequate. Furthermore, time-reversal of the scattered wave is highly challenging since any inaccuracy in the reconstructed beam is amplified by nonlinear propagation. Determining if and to what degree the propagation of light can still be controlled in scattering media with nonlinearity is both a fundamental and practical question, with implications for techniques employing short pulses of light, including nonlinear imaging, laser microsurgery and nonlinear photodynamic therapy \cite{Yelin1999,Evans2008,Saar2010a,Hoy2014,Heisterkamp2002,Bhawalker1997}, focusing of light through multimode fibers \cite{Bianchi2012,Cizmar2011,Wright2015}, as well as for applications employing high power beams, which experience thermal nonlinearity in a variety of samples \cite{Motamedi1988,Lin1996}.

\begin{figure*}
\centerline{\includegraphics[width=15cm]{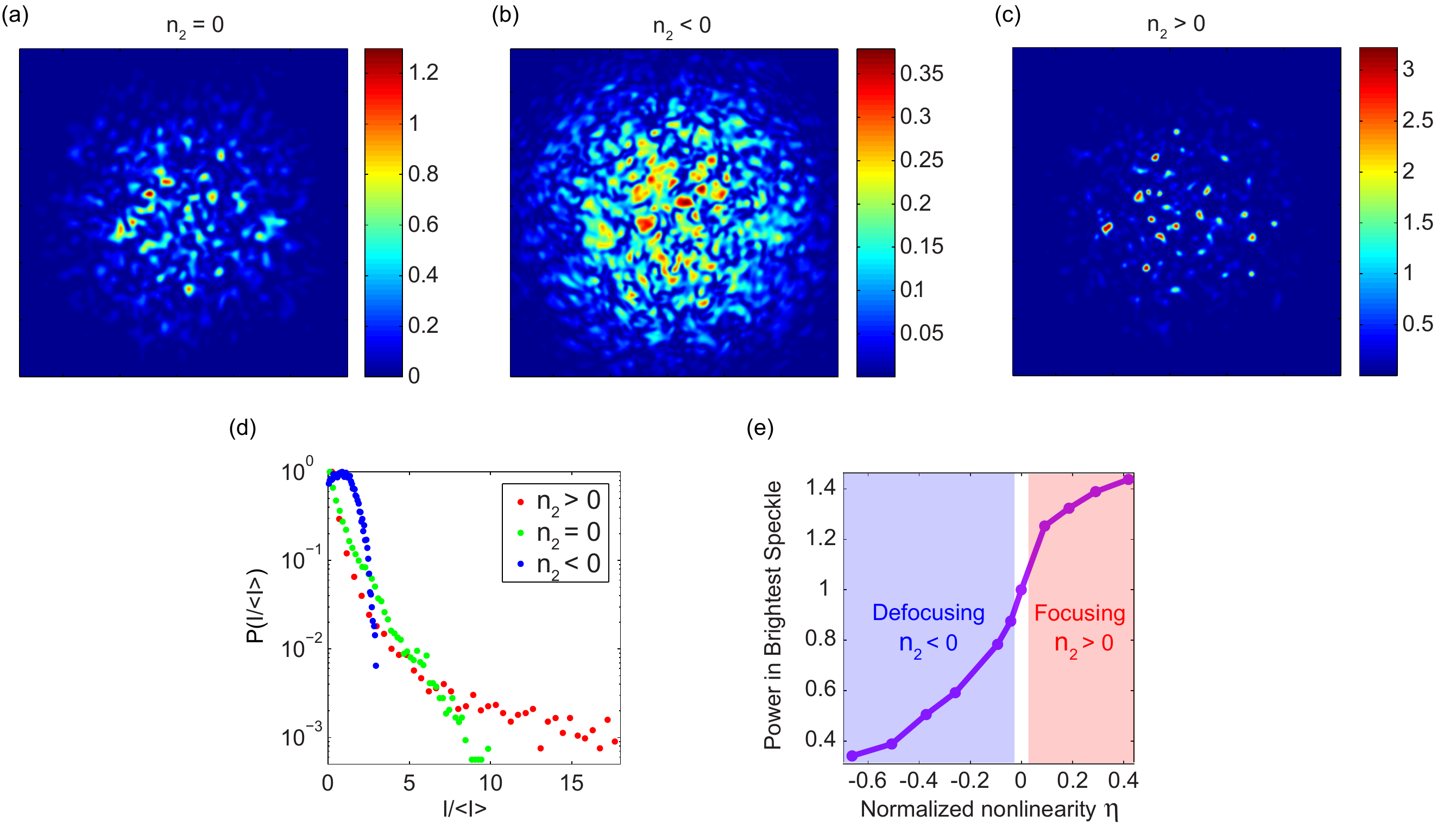}}
\caption{\footnotesize Simulation results of 2D speckle fields after propagation in (a) a linear medium (b) a nonlinear defocusing medium and (c) a nonlinear focusing medium. (d) The probability distribution of the intensities of the speckle fields shown in a-c. (e) The power contained in the brightest speckle as a function of normalized nonlinearity, $\eta\equiv I_{\text{in}}k_0n_2L$ (\textit{without} wavefront shaping). For $\eta>0$ (and $n_2>0$) a larger fraction of the total power is contained within a smaller number of bright speckles, whereas for $\eta<0$ the total power is spread out more evenly among the speckles in the field, compared to the linear case.}
\label{SpecPropag}
\end{figure*}

In this work we study the focusing of light through a scattering medium in the presence of nonlinearity using adaptive optimization of the input wavefront, and show that focusing nonlinearity in fact serves to increase the fraction of total power that the focus contains. Since adaptive optimization iteratively adjusts the solution rather than relying on calibration measurements, it is well-suited for nonlinear media. To elucidate the roles of different elements in the scheme, we studied wavefront shaping in a simple model system consisting of a thin scattering layer followed by a non-scattering nonlinear layer, both numerically and experimentally. We show numerically that this simplified system produces the same qualitative results as a system in which the scattering and nonlinearity are distributed throughout the medium.

We note that speckles propagating through nonlinear media often change their width and shape due to self-focusing or defocusing effects \cite{Bromberg2010}. This change affects the maximum intensity of the speckle without affecting the power it contains. As a result, examining the enhancement of a given speckle by wavefront shaping, defined as the ratio of the maximum intensity of the enhanced speckle to the mean intensity of the unoptimized speckle field, can be ambiguous in nonlinear media. Therefore throughout this work we will analyze the \textit{focused power fraction} achieved by wavefront shaping, which is the ratio of the power contained in the enhanced speckle, or focus, to the total power contained in the speckle field.

Naively, one may expect that the combination of wavefront shaping and nonlinearity can be thought of as two separate effects operating on the speckle pattern sequentially. That is, the total effect can be predicted by combining the increase in power in the focus expected due to wavefront shaping in a linear medium, with the expected self-focusing or self-defocusing of a single speckle caused by the nonlinear medium. Consequently the focus will become narrower (broader) in the presence of focusing (defocusing) nonlinearity due to the self-action of the enhanced speckle, but the fraction of incident power contained in this speckle will not change compare to the linear case. In fact, our results indicate that the combination of the two effects leads to a modified collective behavior, which, for focusing nonlinearity, allows for a greater fraction of the incident power to be controlled and focused. The opposite is true for defocusing nonlinearity, where the fraction of incident power that can be controlled is smaller than for a linear medium. To explain this result we will first discuss how nonlinearity modifies the properties of speckle fields in general, before we consider the added effect of wavefront shaping.

\subsection{Nonlinear propagation of 2D speckle fields}
The propagation of a two-dimensional speckle field inside a medium with Kerr-type nonlinearity is described by the nonlinear wave equation \cite{Boyd2008b}
\begin{equation}\label{NLwaveEq}
i \frac{\partial E}{\partial z}+\frac{1}{2k_0n_0}\left[\frac{\partial^2}{\partial x^2} + \frac{\partial^2}{\partial y^2}\right]E+k_0n_2|E|^2E=0	 \end{equation}
Where $n = n_0 + n_2 I$ is the total refractive index of the medium, $I = |E|^2$, $k_0 = \frac{2\pi}{\lambda}$ and $\lambda$ is the wavelength in vacuum. The last term in the left-hand side of Eq.~\ref{NLwaveEq} causes variations in the refractive index that are proportional to the local intensity of the light propagating through it. The sign of $n_2$ determines whether the nonlinearity experienced by the propagating field will cause it to focus (for $n_2 > 0$) or defocus (for $n_2 < 0$). The variations in the refractive index will cause the different field modes to interact as they propagate, modifying the speckle statistics \cite{Bromberg2010,Derevyanko2012a,Sun2012,Araujo1998} (and can even lead to instabilities in strongly scattering media \cite{Skipetrov2000}). This is demonstrated in the simulation results presented in Fig.~\ref{SpecPropag}, which shows speckle fields after propagation through a linear medium, (Fig.~\ref{SpecPropag}a), a medium with defocusing nonlinearity (Fig.~\ref{SpecPropag}b), and a medium with focusing nonlinearity (Fig.~\ref{SpecPropag}c). The simulation follows Eq.~\ref{NLwaveEq}, yet in order to prevent catastrophic collapse of the speckles for the nonlinear focusing case, a higher-order defocusing term was added to introduce saturation (see methods). Fig.~\ref{SpecPropag}d presents the probability distribution of the intensities of the three speckle fields shown in Fig.~\ref{SpecPropag}a-c. We notice that focusing nonlinearity suppresses the probability of low intensities and enhances the probability of high intensities, resulting in a long-tailed intensity probability distribution (red data in Fig.~\ref{SpecPropag}d) compared to the linear case (green data in Fig.~\ref{SpecPropag}d), whereas the opposite holds true for defocusing nonlinearity (blue data in Fig.~\ref{SpecPropag}d).

Notably, we found that the change in speckle statistics is caused in part by a redistribution of the total power between the speckles and not just by the narrowing (broadening) of the speckles themselves, which would result only in higher (lower) peak intensities. Fig.~\ref{SpecPropag}e depicts the variation of the \textit{total} power contained in the brightest speckle in the speckle field after propagation through media with varying amounts of nonlinearity, averaged over many realizations (\emph{without} wavefront shaping). The nonlinearity values are given as a normalized quantity, $\eta\equiv I_{\text{in}}k_0n_2L$, where $I_{\text{in}}$ is the intensity of the incident field and $L$ is the total propagation distance. $\eta$ therefore represents the total nonlinear phase accumulated during propagation in the medium. The two extreme points in Fig.~\ref{SpecPropag}e, namely $\eta=-0.66$ and $\eta=0.42$, match the values of nonlinearity used to create Fig.~\ref{SpecPropag}b and Fig.~\ref{SpecPropag}c, respectively. The power contained in the speckle was estimated by fitting it to a 2D gaussian function and integrating the result. The values of power are given relative to the power contained in the brightest speckle in a linear medium ($\eta=0$). We see that the power contained in the speckle declines with $\eta$ for defocusing nonlinearity, and grows with $\eta$ for focusing nonlinearity. Hence, focusing nonlinearity redistributes the total power among the speckles such that a smaller number of bright speckles holds a larger fraction of the total power of the field. Defocusing nonlinearity redistributes the total power such that it is spread out more evenly, among many speckles. We expect that this redistribution of power may influence the fraction of incident power that can be controlled and focused using wavefront shaping.

\section{Simulation results}
\begin{figure}[htbp]
\centerline{\includegraphics[width=\columnwidth]{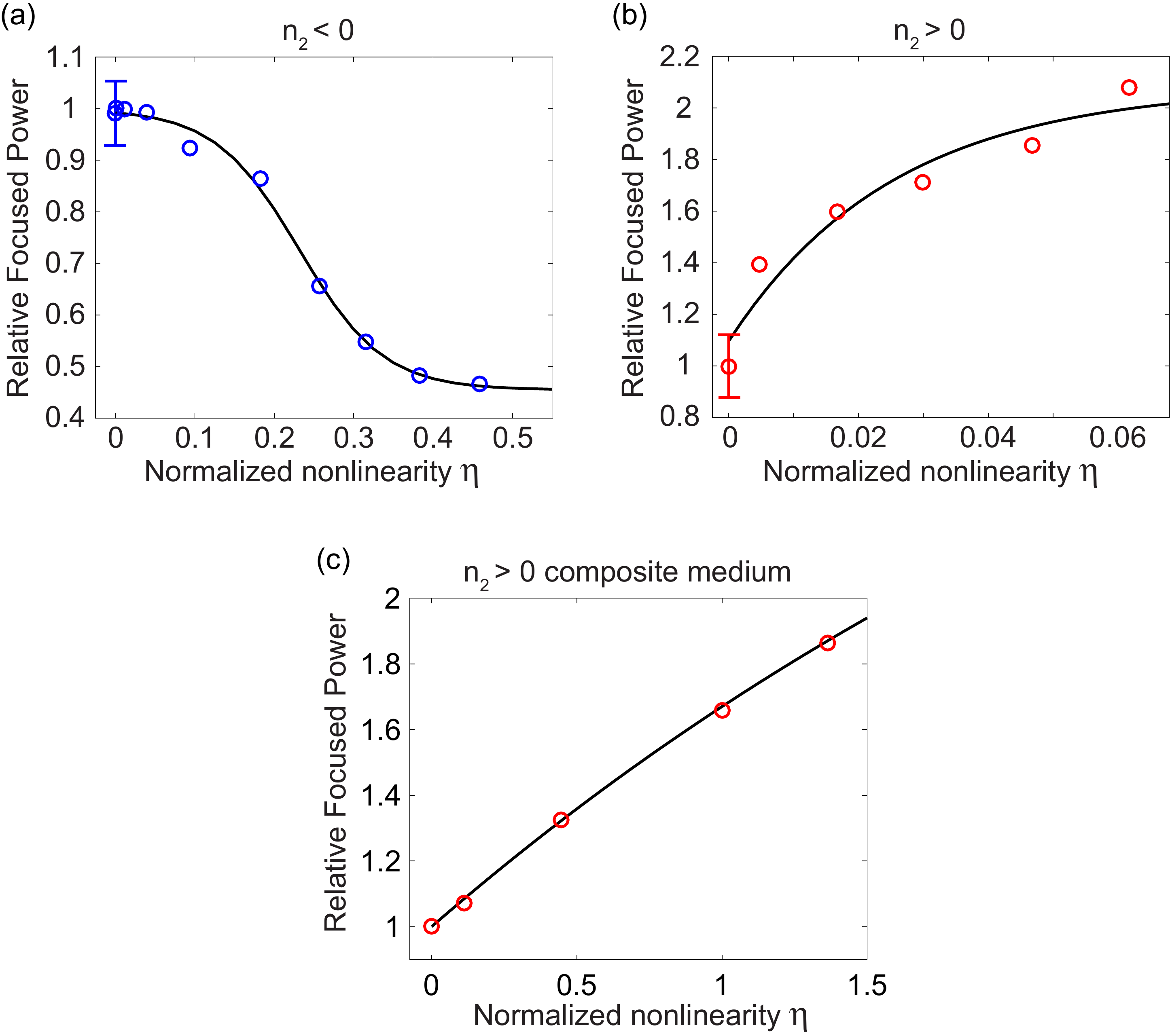}}
\caption{\footnotesize Simulation results: (a) The fraction of the total incident power contained in the obtained focus after optimization, in a medium with defocusing nonlinearity, as a function of normalized nonlinearity $\eta\equiv I_{\text{in}}k_0n_2L$. The focused power fraction declines with nonlinearity strength. (b) Similarly, in a medium with focusing nonlinearity, where the focused power grows with nonlinearity strength. (c) Similarly, in a composite scattering nonlinear focusing medium, consisting of multiple alternating scattering and nonlinear layers. The trend in c is similar to that in b, showing a focused power fraction that grows with nonlinearity strength, though the trend shows less saturation and the $\eta$ values are larger (see SM 1). All focused power fractions were calculated relative to those achieved in a linear scattering medium ($\eta=0$), and all data sets were fitted (black line) to simple functions of exponential form, as a guide to the eye (see methods).}
\label{SimResults}
\end{figure}

Our prediction was tested by performing simulations of wavefront shaping of the field incident upon a forward-scattering layer (diffuser) followed by a nonlinear layer. Between the scattering and nonlinear layers the field was allowed free-space propagation so that the speckle field entering the nonlinear layer was fully developed. The spatial phase of the field incident upon the diffuser was optimized adaptively in order to enhance a single speckle at the output of the nonlinear medium (the configuration of the simulation follows the experimental setup, depicted in Fig.~\ref{Setup} below). This optimization was repeated for different input powers, corresponding to different nonlinearity strengths. The results of this simulation are shown in Fig.~\ref{SimResults}. The data was fitted (black lines) by simple functions of exponential form as a guide to the eye (see methods). We can see that the fraction of power that can be controlled and focused indeed decreases (blue circles in Fig.~\ref{SimResults}a) or increases (red circles in Fig.~\ref{SimResults}b) significantly as a function of nonlinearity strength. This result shows that wavefront shaping causes a dynamic redistribution of power between the speckles. In particular, in a focusing nonlinear medium there is a positive feedback effect: as power is being focused into the region of the chosen speckle by wavefront shaping $\eta$ increases, and the power becomes redistributed such that a smaller amount of the speckles holds a larger fraction of the total power. This allows for more power to be controlled and focused by wavefront shaping, which increases $\eta$, and so on. The positive feedback leads to a focus with a larger fraction of the total power, even for $\eta$ values that are relatively low for the unshaped field. For example, from Fig.~\ref{SimResults}b we can see that the power obtained in the enhanced speckle (the focus) is twice as large for $\eta=0.062$ than for the linear case. Yet from Fig.~\ref{SpecPropag}e we see that for $\eta\simeq0.06$ the distribution of power between the speckles in the initial unshaped field is approximately the same as for the linear case. In a defocusing nonlinear medium, however, as more power is focused into a certain region, it spreads out more evenly between a large number of speckles, hindering the ability to focus power to a chosen speckle at the output. The result is a focus that contains a smaller fraction of the total power than for a linear medium.

Moreover, we propose that our simplified model consisting of a scattering layer followed by a nonlinear layer exhibits the same qualitative behavior as a scattering nonlinear medium, in which scattering and nonlinearity are homogenously distributed. In order to test this assumption, we performed simulations of focusing coherent light through a composite medium, consisting of multiple alternating scattering and nonlinear layers \cite{Wang2017}. The configuration used for the multi-layer simulation is identical to that of the single-layer simulation except for the modification of the medium, and that there is no free-space propagation between the scattering and nonlinear layers. A sketch of the configuration used is shown in Fig.~S1. The results of the composite-medium simulation are presented in Fig.~\ref{SimResults}c. The focused power fraction as a function of normalized nonlinearity $\eta\equiv I_{\text{in}}k_0n_2L$ is represented by red circles. $L$ in this case is the total length of all nonlinear layers. The focused power fractions are normalized to that obtained in a linear multi-layer medium ($\eta=0$). The results follow the same trend as those of the single-layer simulation (Fig.~\ref{SimResults}a), showing focused power fractions that grow by approximately a factor of 2 as focusing nonlinearity is increased (the $\eta$ values are larger for this simulation due to the removal of the free-space propagation, see SM 1). Therefore we may conclude that the trends we observe with our simplified model, consisting of a single scattering layer followed by a single nonlinear layer, are general and hold also for a system with scattering and nonlinearity distributed throughout the medium.

\begin{figure*}[htbp]
\centerline{\includegraphics[width=16cm]{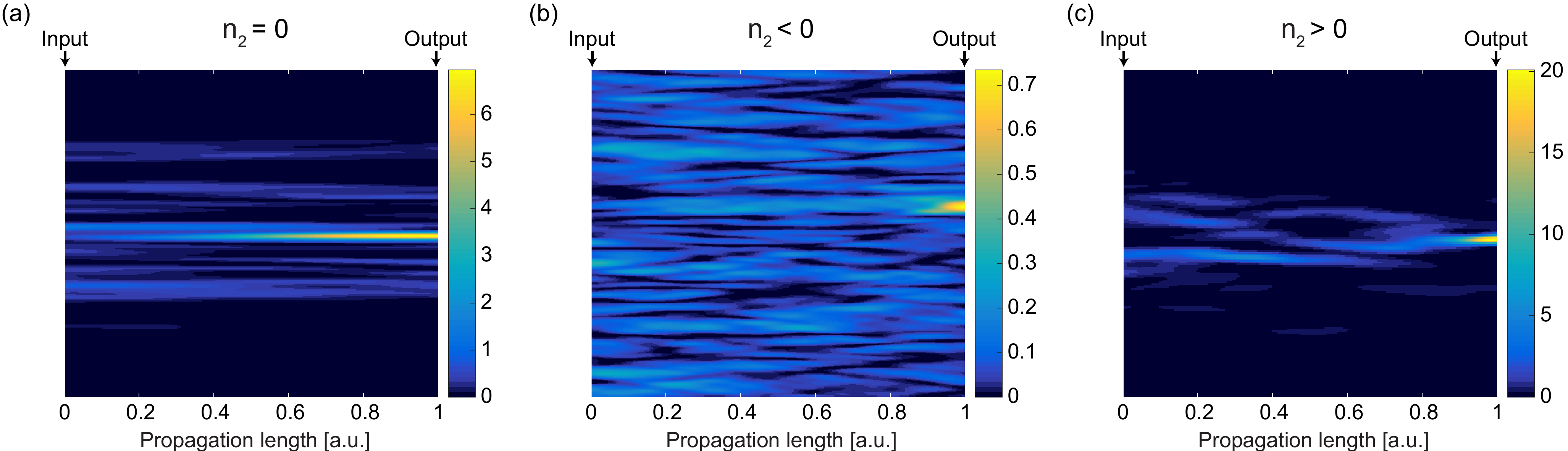}}
\caption{\footnotesize Simulation results: A 1D cut through the propagation of a 2D speckle field, optimized to focus at the output of a layer (at propagation length = 1) with (a) no nonlinearity (b) defocusing nonlinearity and (c) focusing nonlinearity. Tighter axial confinement (along the direction of propagation) of the focus is obtained for both the defocusing and focusing nonlinear cases than for the linear case. For focusing nonlinearity, the trajectories of neighboring bright speckles are altered by wavefront shaping such that they merge into the enhanced speckle.}
\label{OptPropag}
\end{figure*}

The mechanism responsible for the increase in the focused power can be explored by examining the propagation of the optimized field through the nonlinear layer, in the single-layer simulation. Fig.~\ref{OptPropag} shows a one-dimensional cut through the simulated two-dimensional speckle field after wavefront optimization, as it propagates from the input facet of the nonlinear layer (propagation length = 0) towards its output (propagation length = 1), where a focus is created. Since the three fields were optimized to focus after propagation in layers with different nonlinearity types, the optimal SLM phase is different in all three cases and thus also the speckle field entering the nonlinear layer (see Fig.~S2 for a version of Fig.~\ref{OptPropag} without wavefront shaping). In both nonlinear cases the propagation of the speckle field is dynamic, with bright speckles changing quickly into dark and vice versa, whereas in the linear case the propagation is quite static. Accordingly in the linear case, the wavefront optimized to create a focus at the output already contains a bright speckle in that location at the input, which develops to half its final intensity halfway through the medium (Fig.~\ref{OptPropag}a). In the nonlinear cases, both focusing (Fig.~\ref{OptPropag}b) and defocusing (Fig.~\ref{OptPropag}c), the optimized wavefront does not initially contain a bright speckle at the focus location. Furthermore, the enhanced speckle reaches half of its maximal intensity only close to the output, at propagation length $\simeq$ 0.9, creating a tightly confined focus along the propagation direction. In addition, one can observe in Fig.~\ref{OptPropag}c that several bright speckles follow trajectories leading into the enhanced speckle as they propagate from the input to the output, almost as we would expect when focusing a non-speckled field with a lens (or with self-focusing). Thus, in a focusing nonlinear medium, wavefront control can be used to direct energy from neighboring bright speckles into the enhanced speckle.

\section{Experimental results}
\subsection{Focusing light through scattering media in the presence of nonlinearity}
Our experimental scheme is presented in Fig.~\ref{Setup} below. Coherent light passed through an SLM, where it acquired a spatially dependent phase. The shaped light was imaged with a telescope onto a thin scattering medium, and after some free-space propagation entered a nonlinear medium where it propagated through 5~cm. The output facet of the nonlinear medium was imaged onto a CCD camera. An iterative algorithm searched for the optimal SLM phase, which maximized the power of a chosen speckle at the output of the nonlinear medium. For the focusing nonlinearity experiment, the light source was a pulsed femtosecond laser and the nonlinear medium used was ethanol, which exhibits Kerr effect. For the defocusing nonlinearity experiment, the light source was a CW laser and the nonlinear medium used was a weakly-absorbing dye diluted in ethanol, which exhibits defocusing (thermal) nonlinearity \cite{Bromberg2010,Derevyanko2012a} (for more details see methods).

\begin{figure}[htbp]
\centerline{\includegraphics[width=\columnwidth]{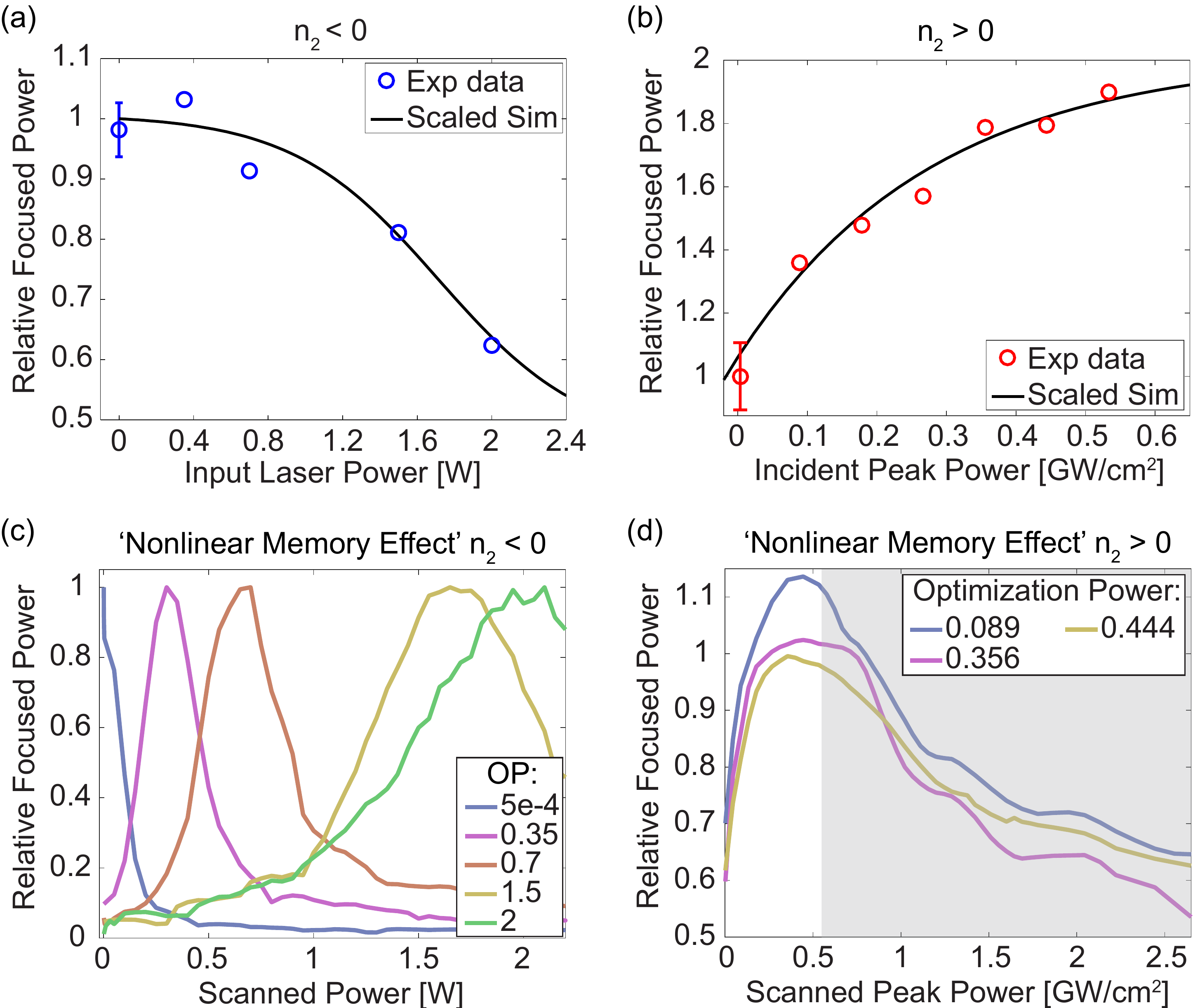}}
\caption{\footnotesize Experimental Results: focusing light through scattering media in the presence of nonlinearity. (a) The fraction of the total incident power contained in the obtained focus after optimization in a defocusing nonlinear medium, as a function of the input laser power (blue circles). The scaled simulation results are plotted for comparison (black line). (b) The focused power fraction obtained in a focusing medium, as a function of the peak power incident on the diffuser (red circles), and the scaled simulation (black line). The fraction of light that can be controlled and focused rises (declines) with nonlinearity strength in a focusing (defocusing) medium, as predicted by the simulation. (c) A measurement of the `nonlinear memory effect' in the defocusing medium, i.e. the degree to which the optimal SLM phase can be used to obtain focusing as the nonlinearity strength of the medium is varied, without reoptimization. The different colored lines represent the results for the different optimization powers shown in the legend. (d) A measurement of the `nonlinear memory effect' in the focusing medium. The white region is the range of peak powers used in (b). While for defocusing media the optimal SLM phase performs best for the optimization power and gradually declines in efficiency, for focusing media the optimal SLM phase is very robust and achieves a focus of similar quality for peak power $\sim$5-10 times larger than the optimization power.}
\label{ExpResults}
\end{figure}

The experimental results of the obtained focused power fractions at different laser powers are presented in Fig.~\ref{ExpResults}. Fig.~\ref{ExpResults}a shows the fraction of focused power in a defocusing nonlinear medium (blue circles) as a function of the input laser power, and the scaled simulation results for comparison (black line). Fig.~\ref{ExpResults}b shows the fraction of focused power in a focusing nonlinear medium (red circles) as a function of the peak power incident on the diffuser, and the scaled simulation results for comparison (black line). The focused power fractions in both figures are relative to the linear case. In both experiments, the results follow the trend of the simulation, either increasing or decreasing with nonlinearity strength. The experimental results thus verify the prediction that moderate nonlinearity significantly alters the fraction of power that can be controlled and focused through a scattering medium, and that the focused power fraction can be substantially increased in the presence of mild focusing nonlinearity. The range of powers shown in Fig.~\ref{ExpResults}a corresponds to the low nonlinearity range of the simulation ($\eta\lesssim0.28$ in Fig.~\ref{SimResults}a), while the range of powers shown in Fig.~\ref{ExpResults}b approximately corresponds to the full nonlinearity range of the simulation (Fig.~\ref{SimResults}b). The peak powers used in the focusing nonlinear experiment are over an order of magnitude lower than those typically used at the focus during two-photon microscopy, which are approximately in the $\sim$10~GW/cm$^2$ range. Therefore, even taking into account the $\sim$20-fold increase in $\eta$ values predicted by the multi-layer simulation, we expect significant changes in the focused power fraction to occur in scattering samples with high water content during nonlinear microscopy measurements. We note that the nonlinearity of the defocusing medium used is fundamentally nonlocal, whereas the simulated defocusing nonlinearity is local. Yet the agreement between the experimental and simulation results shows that the nonlocality does not significantly alter the results of our experiment.

\begin{figure}[htbp]
\centerline{\includegraphics[scale = 0.125, bb=0 0 2000 1500]{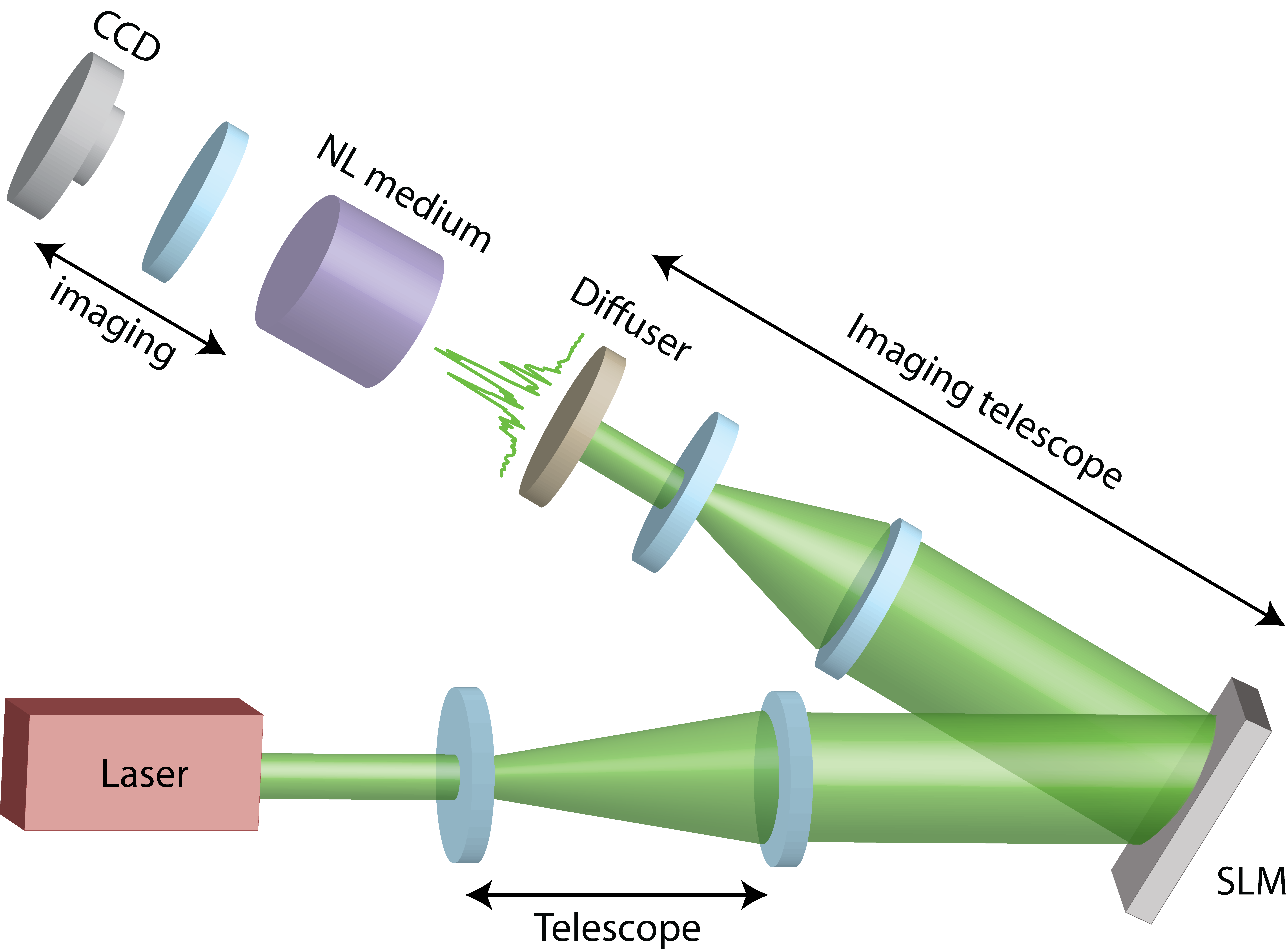}}
\caption{\footnotesize The experimental setup. The wavefront of either a CW (for defocusing nonlinearity) or a pulsed (for focusing nonlinearity) laser was shaped by a two-dimensional spatial light modulator (SLM) and then imaged with a telescope onto a thin diffuser. The generated speckle pattern propagated through a nonlinear medium (NL medium) and the output facet of the nonlinear medium was imaged onto a CCD camera.}
\label{Setup}
\end{figure}

\subsection{The `nonlinear memory effect'}
Another interesting aspect of focusing of light through nonlinear scattering media is the robustness of the optimization to fluctuations or changes in the nonlinearity strength. Such fluctuations can be caused, for example, by fluctuations in the intensity of the laser. To address this question experimentally, after finding the optimal SLM phase for a certain laser power, we measured the fraction of focused power obtained with this SLM phase for other laser powers. This measurement was performed for all optimization powers. In a sense, this is a measure of the `nonlinear memory effect' \cite{Freund1988} of the scattering medium, i.e. the degree to which the optimal SLM phase can be used to obtain a focus as the nonlinearity strength of the medium is varied, without reoptimization.

The results are presented in Fig.~\ref{ExpResults}c for defocusing nonlinearity and Fig.~\ref{ExpResults}d for focusing nonlinearity. The different colors represent measurements performed for different optimization powers, shown in the legend. The results for the two cases are qualitatively different. In the defocusing case, we see a clear drop in the focused power on either side of the optimization power, and the drop is roughly symmetrical between the two sides. Furthermore, for larger optimization powers, corresponding to larger nonlinearity values (and lower focused power fractions) the widths of the curves grow. Therefore, the optimization is more robust and the optimal SLM phase can be used to obtain a focus of similar quality for larger variations in the nonlinearity strength compared to lower nonlinearity values. In contrast, in the focusing nonlinear case the curves for all optimization powers are quite similar (the curves of the other four optimization powers were omitted from the figure for clarity). The optimization is very robust and can be used to obtain a focus of similar quality all throughout the nonlinearity range examined in the experiment (0.1~GW/cm$^2 <$ laser peak power $<$ 0.5~GW/cm$^2$) and beyond it, up to $\sim$1~GW/cm$^2$. For powers below 0.1~GW/cm$^2$ the focused power fraction drops rapidly, whereas for power above 1~GW/cm$^2$ the focused power fraction declines more slowly, reaching half of the original value only around $\sim$5-10 times the optimization powers.

\section{Discussion}
We have demonstrated, both numerically and experimentally, focusing of light through a scattering layer followed by a nonlinear layer, and shown that nonlinearity significantly modifies the fraction of power that can be channeled to the focus. In particular, the presence of moderate focusing nonlinearity has been shown to cause about a two-fold increase in the total power in the obtained focus compared to a linear medium. Conversely, the presence of moderate defocusing nonlinearity has been shown to cause approximately a two-fold decrease in the focused power fraction. Our simulations show that these findings are valid also when the scattering and nonlinearity are homogeneously distributed throughout the medium.

The results presented in this work suggest the favorability of focusing and imaging through scatterers with short pulses of light, which experience mild Kerr self-focusing in a variety of samples, including biologically-relevant samples \cite{Rockwell1993,Potma2001,Hunt2007}. Previous works have shown that pulsed fields that propagate through scattering layers maintain temporal coherence long after the spatial coherence is lost \cite{Katz2011a}, and that nonlinear signals can be utilized to focus light noninvasively \cite{Katz2014}. Our current results show that the basic premises of wavefront shaping indeed extend into the nonlinear regime, and we expect that such noninvasive nonlinear techniques might be even enhanced by the nonlinear response of the specimen.

\section{Methods}
\subsection{Single-layer simulation implementation}
Nonlinear propagation was simulated using the split-step method. Time-domain was not addressed specifically in the simulation, as this would result in impractical run times. The diffuser was modeled as a single layer, with no thickness, of randomly distributed phase features, $\phi$, in the range of $-\pi<\phi<\pi$. The scattering is therefore only in the forward direction.

Defocusing nonlinearity was modeled as Kerr-type, with the nonlinear operator $\hat{N}=-k_0n_2|E|^2$. The beam illuminated $\sim$110x110 features on the modeled diffuser, which resulted in a speckle field that contained several thousands of speckles. The SLM was modeled with 15x15 pixels. The linear ($\eta=0$) enhancement value and focused power fraction were 300 and 0.022, respectively. However, in the simulation \textit{without} wavefront shaping shown in Fig.~\ref{SpecPropag}b, the implementation of the defocusing simulation matched that of the focusing simulation exactly, for the purpose of comparison. Therefore the beam illuminated $\sim$30x30 features on the modeled diffuser, which resulted in a speckle field that contained several hundreds of speckles.

Focusing nonlinearity was modeled with a higher-order defocusing term, $\hat{N}=k_0(n_2|E|^2-n_4|E|^4)$, where $n_4 \ll n_2$, for both the simulations with wavefront shaping and without. The higher-order term introduces saturation of the self-focusing process, as is often observed in various Kerr samples \cite{Bree2011}, and prevents the collapse of the speckles to sizes below the resolution of the simulation. The beam illuminated $\sim$30x30 features on the modeled diffuser, which resulted in a speckle field that contained several hundreds of speckles. The SLM was modeled with 8x8 pixels. The simulation parameters were modified for the focusing case because of its increased run time (due to stronger nonlinearity effects that require more propagation steps). The simulated window size was therefore reduced in order to compensate for the increase. In order to equate the nonlinearity scales of the two simulations, the $\eta$ values for the defocusing simulation were divided by a factor of 2. The linear ($\eta=0$) enhancement value and focused power fraction were 77 and 0.0413, respectively.

For both simulations the speckles entering the media were fully developed, and $\eta$ was varied by varying $I_{\text{in}}$. An arbitrary speckle at the output of the nonlinear layer was chosen and optimized using an iterative genetic algorithm. The functions used for fitting the simulation results in Fig.~\ref{SimResults} were: (a) $f(x) = \frac{a}{\exp(bx - c)+1}+ d$ and (b) $f(x) = a \exp(-bx) + c$. For details on the implementation of the multi-layer simulation, see SM 1.

\subsection{Experimental implementation}
In the defocusing nonlinearity experiment, light from a 532~nm CW laser was expanded with a 6x magnifying telescope and reflected off a two-dimensional liquid-crystal SLM (Hamamatsu LCOS X10468-01). The phase-shaped wavefront was imaged with a 6x demagnifying telescope onto a holographic diffuser (5$^{\circ}$ diffusion FWHM) and propagated $\sim$5~cm in air before it entered the nonlinear medium. The transmission of the diffuser was $\sim$90$\%$. The output facet of the nonlinear medium was then imaged onto a CCD camera (Andor Luca S). An arbitrary speckle was chosen and optimized using an iterative genetic algorithm. The nonlinear medium was LDS 751 dye dissolved in ethanol contained in a 5~cm path length cylindrical cuvette. The solution weakly absorbed the light due to the low concentration of the dye and exhibited thermal nonlinearity due to the ethanol. Of the total laser power, $\sim$70$\%$ entered the cuvette and $\sim$50$\%$ of that was absorbed in it. The genetic algorithm was terminated for all measurements after 1550 generations, a number of generations that showed sufficient convergence for all runs. The enhancement value for the linear run (0.5~mW input power) was 650. An accurate absolute measure of the focused power fraction is more challenging to obtain in the experimental implementation, due to the finite aperture of the CCD, and therefore only relative focused power fractions were calculated.

In the focusing nonlinearity experiment, 35~fsec pulses from an amplified femtosecond laser (bandwidth of $\sim$30~nm, centered at 800~nm) were significantly attenuated using a waveplate and a polarizing beam splitter to obtain the desired peak powers. The beam was directly reflected off a broadband two-dimensional SLM (Hamamatsu LCOS X10468-02). The phase-shaped wavefront was imaged with a 3.33x demagnifying telescope onto a holographic diffuser (1$^{\circ}$ diffusion FWHM) and propagated $\sim$15~cm in air before it entered the nonlinear medium. The output facet of the nonlinear medium was then imaged onto the CCD camera, where a chosen speckle was optimized using an iterative genetic algorithm. The nonlinear medium was neat ethanol contained in a 5~cm path length cylindrical cuvette. The peak powers provided in Fig.~\ref{ExpResults}b and d are those incident on the diffuser. The genetic algorithm was terminated for all measurements after $\sim$1300 generations, which showed sufficient convergence for all runs. The enhancement value for the linear run (3.6e-3~GW/cm$^2$ incident peak power) was 300.

\subsection{Evaluation of Focused Power}
The amount of power contained in the focused speckle, in both the simulations and the defocusing nonlinearity experiment, was calculated by fitting it with a two-dimensional gaussian function with a different width in each dimension and integrating the fitted expression. The gaussian fit was not accurate enough in the focusing nonlinearity experiment and therefore the energy in the focused speckle was calculated simply by integrated over the speckle area. The error bars for the $\eta = 0$ data point for Fig.~\ref{SimResults} and Fig.~\ref{ExpResults} are presented as an example of the typical error in these plots. The error values were calculated by applying the focusing procedure 20 times using the same parameters and calculating the standard deviation of the obtained focused power fractions for the simulations and defocusing experiment. For the focusing experiment the errors were calculated with 9 runs.

\section*{Acknowledgments}
The authors thank R. Fischer, M. Segev, H.~H. Sheinfux and D. Gilboa for helpful discussions.

\footnotesize{
\bibliographystyle{ieeetr}

\begin{thebibliography}{10}

\bibitem{Vellekoop2007a}
I.~M. Vellekoop and A.~P. Mosk, ``{Focusing coherent light through opaque
  strongly scattering media},'' {\em Opt. Lett.}, vol.~32, pp.~2309--2311,
  2007.

\bibitem{Cizmar2010}
T.~Cizmar, M.~Mazilu, and K.~Dholakia, ``{In situ wavefront correction and its
  application to micromanipulation},'' {\em Nat. Photonics}, vol.~4,
  pp.~388--394, 2010.

\bibitem{Vellekoop2010}
I.~M. Vellekoop, A.~Lagendijk, and A.~P. Mosk, ``{Exploiting disorder for
  perfect focusing},'' {\em Nat. Photonics}, vol.~4, no.~February,
  pp.~320--322, 2010.

\bibitem{Popoff2010}
S.~Popoff, G.~Lerosey, M.~Fink, A.~C. Boccara, and S.~Gigan, ``{Image
  transmission through an opaque material},'' {\em Nat. Commun.}, vol.~1,
  pp.~1--5, jan 2010.

\bibitem{vanPutten2011}
E.~G. van Putten, D.~Akbulut, J.~Bertolotti, W.~L. Vos, A.~Lagendijk, and A.~P.
  Mosk, ``{Scattering lens resolves sub-100 nm structures with visible
  light},'' {\em Phys. Rev. Lett.}, vol.~106, p.~193905, may 2011.

\bibitem{Xu2011}
X.~Xu, H.~Liu, and L.~V. Wang, ``{Time-reversed ultrasonically encoded optical
  focusing into scattering media},'' {\em Nat. Photonics}, vol.~5,
  pp.~154--157, 2011.

\bibitem{Katz2011a}
O.~Katz, E.~Small, Y.~Bromberg, and Y.~Silberberg, ``{Focusing and compression
  of ultrashort pulses through scattering media},'' {\em Nat. Photonics},
  vol.~5, no.~6, pp.~372--377, 2011.

\bibitem{Aulbach2011}
J.~Aulbach, B.~Gjonaj, P.~M. Johnson, A.~P. Mosk, and A.~Lagendijk, ``{Control
  of light transmission through opaque scattering media in space and time},''
  {\em Phys. Rev. Lett.}, vol.~106, no.~March, p.~103901, 2011.

\bibitem{McCabe2011}
D.~J. McCabe, A.~Tajalli, D.~R. Austin, P.~Bondareff, I.~A. Walmsley, S.~Gigan,
  and B.~Chatel, ``{Spatio-temporal focusing of an ultrafast pulse through a
  multiply scattering medium},'' {\em Nat. Commun.}, vol.~2, pp.~1--5, 2011.

\bibitem{Choi2011}
Y.~Choi, T.~D. Yang, C.~Fang-Yen, P.~Kang, K.~J. Lee, R.~R. Dasari, M.~S. Feld,
  and W.~Choi, ``{Overcoming the diffraction limit using multiple light
  scattering in a highly disordered medium},'' {\em Phys. Rev. Lett.},
  vol.~107, p.~023902, jul 2011.

\bibitem{Yaqoob2008}
Z.~Yaqoob, D.~Psaltis, M.~S. Feld, and C.~Yang, ``{Optical phase conjugation
  for turbidity suppression in biological samples},'' {\em Nat. Photonics},
  vol.~2, pp.~110--115, jan 2008.

\bibitem{Wang2012}
Y.~M. Wang, B.~Judkewitz, C.~A. Dimarzio, and C.~Yang, ``{Deep-tissue focal
  fluorescence imaging with digitally time-reversed ultrasound-encoded
  light},'' {\em Nat. Commun.}, vol.~3, no.~May, pp.~1--8, 2012.

\bibitem{Katz2012}
O.~Katz, E.~Small, and Y.~Silberberg, ``{Looking around corners and through
  thin turbid layers in real time with scattered incoherent light},'' {\em Nat.
  Photonics}, vol.~6, pp.~1--5, 2012.

\bibitem{Mosk2012}
A.~P. Mosk, A.~Lagendijk, G.~Lerosey, and M.~Fink, ``{Controlling waves in
  space and time for imaging and focusing in complex media},'' {\em Nat.
  Photonics}, vol.~6, pp.~283--292, 2012.

\bibitem{Barsi2009}
C.~Barsi, W.~Wan, and J.~W. Fleischer, ``{Imaging through nonlinear media using
  digital holography},'' {\em Nat. Photonics}, vol.~3, no.~April, pp.~211--215,
  2009.

\bibitem{Goy2011}
A.~Goy and D.~Psaltis, ``{Digital reverse propagation in focusing Kerr
  media},'' {\em Phys. Rev. A}, vol.~83, p.~031802, 2011.

\bibitem{Yelin1999}
D.~Yelin and Y.~Silberberg, ``{Laser scanning third-harmonic-generation
  microscopy in biology},'' {\em Opt. Express}, vol.~5, pp.~169--175, 1999.

\bibitem{Evans2008}
C.~L. Evans and X.~{Sunney Xie}, ``{Coherent anti-stokes Raman scattering
  microscopy: chemical imaging for biology and medicine.},'' {\em Annu. Rev.
  Anal. Chem.}, vol.~1, pp.~883--909, jan 2008.

\bibitem{Saar2010a}
B.~G. Saar, C.~W. Freudiger, J.~Reichman, C.~M. Stanley, G.~R. Holtom, and
  X.~{Sunney Xie}, ``{Video-rate molecular imaging in vivo with stimulated
  Raman scattering},'' {\em Science}, vol.~330, pp.~1368--1370, 2010.

\bibitem{Hoy2014}
C.~L. Hoy, O.~Ferhanoglu, M.~Yildirim, K.~H. Kim, S.~S. Karajanagi, K.~M.~C.
  Chan, J.~B. Kobler, S.~M. Zeitels, and A.~Ben-yakar, ``{Clinical ultrafast
  laser surgery: recent advances and future directions},'' {\em IEEE J. Quantum
  Electron.}, vol.~20, p.~7100814, 2014.

\bibitem{Heisterkamp2002}
A.~Heisterkamp, T.~Ripken, T.~Mamom, W.~Drommer, H.~Welling, W.~Ertmer, and
  H.~Lubatschowski, ``{Nonlinear side effects of fs pulses inside corneal
  tissue during photodisruption},'' {\em Appl. Phys. B Lasers Opt.}, vol.~74,
  pp.~419--425, apr 2002.

\bibitem{Bhawalker1997}
J.~D. Bhawalker, N.~D. Kumar, C.-F. Zhao, and P.~N. Prasad, ``{Two-photon
  photodynamic therapy},'' {\em J. Clin. Laser Med. Surg.}, vol.~15,
  pp.~201--204, 1997.

\bibitem{Bianchi2012}
S.~Bianchi and R.~{Di Leonardo}, ``{A multi-mode fiber probe for holographic
  micromanipulation and microscopy},'' {\em Lab chip}, vol.~12, pp.~635--639,
  2012.

\bibitem{Cizmar2011}
T.~Cizmar and K.~Dholakia, ``{Shaping the light transmission through a
  multimode optical fibre: complex transformation analysis and applications in
  biophotonics},'' {\em Opt. Express}, vol.~19, pp.~18871--18884, 2011.

\bibitem{Wright2015}
L.~G. Wright, D.~N. Christodoulides, and F.~W. Wise, ``{Controllable
  spatiotemporal nonlinear effects in multimode fibres},'' {\em Nat.
  Photonics}, vol.~9, pp.~306--310, 2015.

\bibitem{Motamedi1988}
M.~Motamedi, A.~J. Welch, W.-f. Cheong, S.~A. Ghaffari, and O.~T. Tan,
  ``{Thermal Lensing in Biologic Medium},'' {\em IEEE J. Quantum Electron.},
  vol.~24, pp.~693--696, 1988.

\bibitem{Lin1996}
W.-C. Lin, M.~Motamedi, and A.~J. Welch, ``{Dynamics of tissue optics during
  laser heating of turbid media},'' {\em Appl. Opt.}, vol.~35, pp.~3413--3420,
  1996.

\bibitem{Bromberg2010}
Y.~Bromberg, Y.~Lahini, E.~Small, and Y.~Silberberg, ``{Hanbury Brown and Twiss
  interferometry with interacting photons},'' {\em Nat. Photonics}, vol.~4,
  pp.~721--726, 2010.

\bibitem{Boyd2008b}
R.~W. Boyd, {\em {Nonlinear optics}}.
\newblock Academic Press, 3rd~ed., 2008.

\bibitem{Derevyanko2012a}
S.~Derevyanko and E.~Small, ``{Nonlinear propagation of an optical speckle
  field},'' {\em Phys. Rev. A}, vol.~053816, no.~85, pp.~1--9, 2012.

\bibitem{Sun2012}
C.~Sun, S.~Jia, C.~Barsi, S.~Rica, A.~Picozzi, and J.~W.~Fleischer,
  ``{Observation of the kinetic condensation of classical waves},'' {\em Nat.
  Phys.}, vol.~8, pp.~470--474, apr 2012.

\bibitem{Araujo1998}
R.~E.~de~Araujo and A.~S.~L.~Gomes,
  ``{Nonlinear optical Kerr coefficients of disordered media},'' {\em Phys.
  Rev. A}, vol.~57, pp.~2037--2040, 1998.

\bibitem{Skipetrov2000}
S.~E. Skipetrov and R.~Maynard, ``{Instabilities of waves in nonlinear disordered media},''
{\em Phys. Rev. Lett.}, vol.~85, pp.~736--739, 2000.

\bibitem{Wang2017}
Y.~Wang, J.~W.~Fleischer, ``{A model of thick scattering media using a series of random diffraction gratings},''
{\em Imaging and Applied Optics 2017} paper JTu5A.

\bibitem{Freund1988}
I.~Freund, M.~Rosenbluh, and S.~Feng, ``{Memory effects in propagation of
  optical waves through disordered media},'' {\em Phys. Rev. Lett.}, vol.~61,
  no.~20, pp.~2328--2332, 1988.

\bibitem{Rockwell1993}
B.~A. Rockwell, W.~P. Roach, M.~E. Rogers, M.~W. Mayo, C.~a. Toth, C.~P. Cain,
  and G.~D. Noojin, ``{Nonlinear refraction in vitreous humor},'' {\em Opt.
  Lett.}, vol.~18, pp.~1792--1794, nov 1993.

\bibitem{Potma2001}
E.~O. Potma, W.~P. {De Boeij}, and D.~A. Wiersma, ``{Femtosecond dynamics of
  intracellular water probed with nonlinear optical Kerr effect
  microspectroscopy},'' {\em Biophys. J.}, vol.~80, pp.~3019--3024, 2001.

\bibitem{Hunt2007}
N.~T. Hunt, L.~Kattner, R.~P. Shanks, and K.~Wynne, ``{The Dynamics of
  Water-Protein Interaction Studied by Ultrafast Optical Kerr-Effect
  Spectroscopy},'' {\em J. Am. Chem. Soc.}, vol.~129, pp.~3168--3172, 2007.

\bibitem{Katz2014}
O.~Katz, E.~Small, Y.~Guan, and Y.~Silberberg, ``{Noninvasive nonlinear
  focusing and imaging through strongly scattering turbid layers},'' {\em
  Optica}, vol.~1, no.~3, pp.~170--174, 2014.

\bibitem{Bree2011}
C.~Bree, A.~Demircan, and G.~Steinmeyer, ``{Saturation of the all-optical Kerr
  effect},'' {\em Phys. Rev. Lett.}, vol.~106, p.~183902, 2011.

\end{thebibliography}

}
\end{document}


\twocolumn[
  \maketitle
\begin{onecolabstract}
\normalsize \textbf{This document provides supplementary information to "Focusing light by wavefront shaping through disorder and nonlinearity". The supplementary information includes two sections. The first section provides details on the implementation of the composite scattering nonlinear medium simulation, the results of which are presented in Fig.~2c of the main text. The second section presents cross-sections through simulated two-dimensional speckle propagation in a nonlinear layer without wavefront shaping, for comparison with Fig.~3 of the main text.}
\vspace{10mm}
\end{onecolabstract}]

\section{Focusing coherent light through multiple scattering and nonlinear layers: simulation details}
\begin{figure}[htbp]
\centering
\includegraphics[scale = 0.12, bb=0 0 2025 1500]{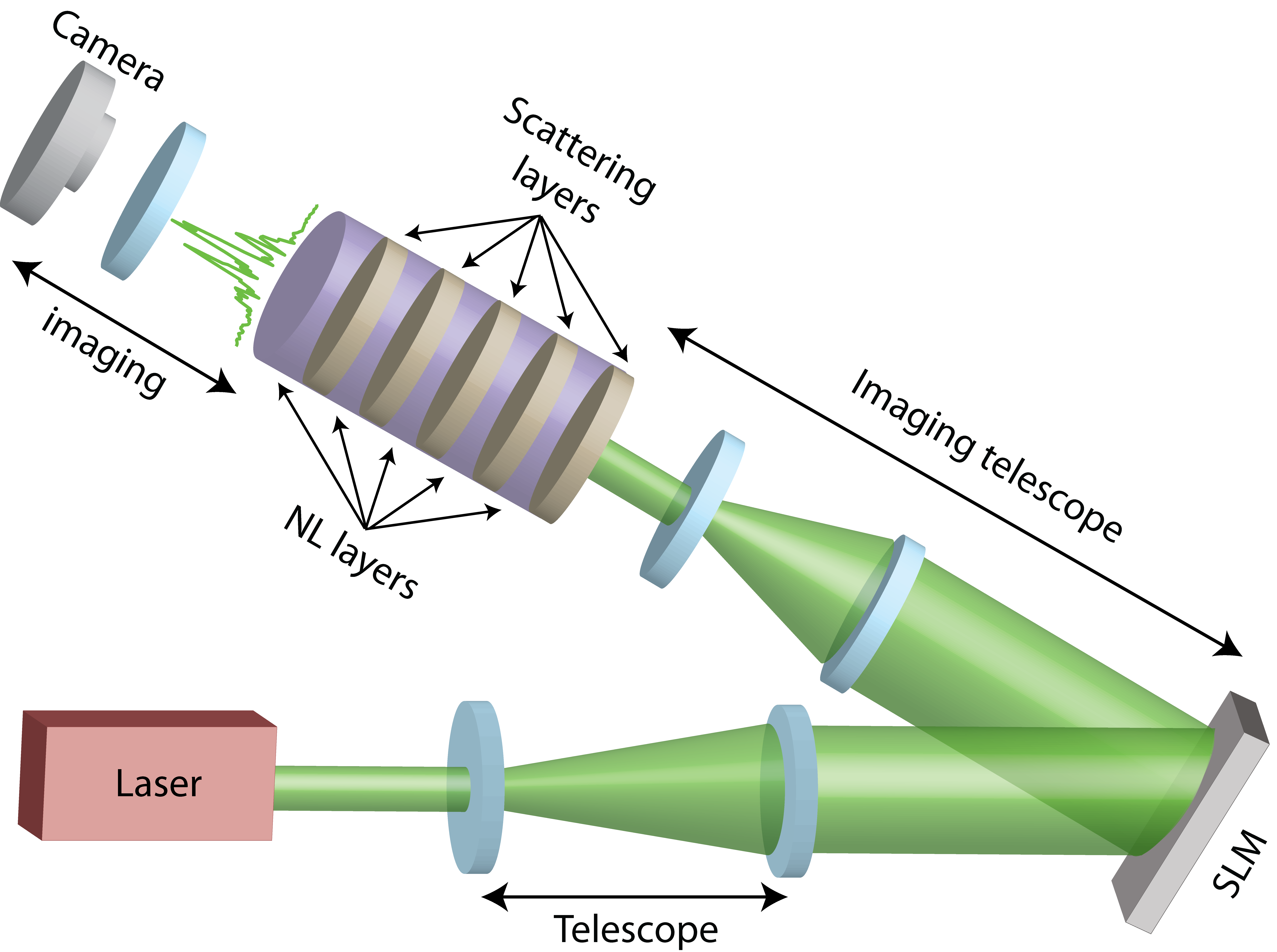}
\caption{Conceptual sketch of the wavefront shaping system used in the composite scattering nonlinear medium simulation. The light in this simulation is scattered by five equidistant thin scattering layers, each followed by a homogeneous nonlinear (NL) layer. The wavefront is shaped with the SLM in order to create a focus at the output facet of the composite medium (or, equivalently, on the camera). The results of this simulation are presented in Fig. 2c of the main text.}
\label{MultDiffSetup}
\end{figure}

In the main text we study focusing of coherent light through a scattering layer followed by a nonlinear layer using wavefront shaping. We propose that this simplified model exhibits the same qualitative behavior as a scattering nonlinear medium, i.e. a medium composed of many successive scattering and nonlinear layers. In order to test this assumption, we performed simulations of focusing coherent light through a composite medium with multiple alternating scattering and nonlinear layers with wavefront shaping. The configuration used for the composite-medium simulation is shown in Fig.~\ref{MultDiffSetup}. The configuration is identical to that used in the single-layer simulation, except for the modification of the medium and that there is no free-space propagation between any of the scattering and nonlinear layers. The composite medium consists of five scattering layers with four nonlinear layers in between them, and a fifth nonlinear layer at the end. The scattering layers were modeled just like in the single-layer simulation, as a sheet of randomly distributed phase features in the range of $-\pi<\phi<\pi$, but the phase features were slightly smoothed to avoid sharp phase jumps. The incident beam illuminated $\sim$30x30 features on the first scattering layer (after smoothing). Despite being multiple scattering, the composite medium is still entirely forward-scattering. The nonlinear layers were modeled with positive nonlinearity ($n_2 > 0$) and were each a fifth of the length of the layer used in the single-layer simulation. The spatial phase of the field incident upon the first scattering layer (or, equivalently, upon the SLM) was optimized adaptively in order to enhance a single speckle at the output of the composite medium, i.e. at the output of last nonlinear layer. This optimization was repeated for different $I_{\text{in}}$ values, corresponding to different nonlinearity strengths. The linear ($\eta=0$) enhancement value and focused power fraction were 79 and 0.0283, respectively.

The results presented in Fig.~2c follow the same trend as Fig.~2b, showing a focused power fraction that increases roughly by a factor of 2 as nonlinearity increases. We note that the $\eta$ values in the multiple-layer simulation are larger than those in the single-layer simulation. This is due to the difference between the configurations that the two simulations are based on. Since in the single-layer simulation the speckle field entering the nonlinear medium is fully developed, it propagates under the influence of local nonlinearity throughout the entire medium. In the multi-layer simulation, however, the scattering layers directly follow the nonlinear layers and therefore the scattered field must propagate through a significant length of the composite medium before the speckles develop and local nonlinearity becomes significant. Despite the increase, the $\eta$ values, or nonlinear phases accumulated, in the multi-layer simulation are still of the order of the values typically used in nonlinear microscopy experiments. Additionally, the trend in 2c shows less saturation than that of 2b. We did not explore larger $\eta$ values, which would likely show this saturation trend, due to limited computing power. Since the nonlinearity values are higher for these points, more steps must be performed during propagation using the split-step method. The computational cost of running a genetic algorithm, which must evaluate the fitness of >30,000 phase patterns using such propagation, resulted in impractical run times.

\section{Simulation results: cross-sections through 2d speckle propagation in nonlinear media without wavefront shaping}
In Fig.~3 we presented one-dimensional cross-sections through the propagation of simulated two-dimensional speckle fields in linear (Fig.~3a), defocusing nonlinear (Fig.~3b) and focusing nonlinear media (Fig.~3c), \textit{after} their wavefront was shaped in order to focus at the output of the medium. For comparison purposes, we present cross-sections through the propagation of the same two-dimensional fields, created by the same diffuser, in the same linear (Fig.~S2a), defocusing nonlinear (Fig.~S2b) and focusing nonlinear media (Fig.~S2c) yet \textit{without} wavefront shaping (hence, the speckle fields entering the nonlinear media are different in Fig.~3 and Fig.~S2). We can see that the linear case looks quite similar with and without wavefront shaping, with the speckle field not changing much during propagation. In both nonlinear cases the speckle field evolves quickly during propagation, bright speckles turning into dark and vice versa, as they do without wavefront shaping. The defocusing nonlinear case is similar with and without wavefront shaping in other aspects as well. However, the focusing nonlinear case without wavefront shaping lacks the effect of multiple bright speckles following trajectories leading into a particular bright speckle, as seen with wavefront shaping.

\begin{figure*}[b]
\centering
\includegraphics[width=18.5cm]{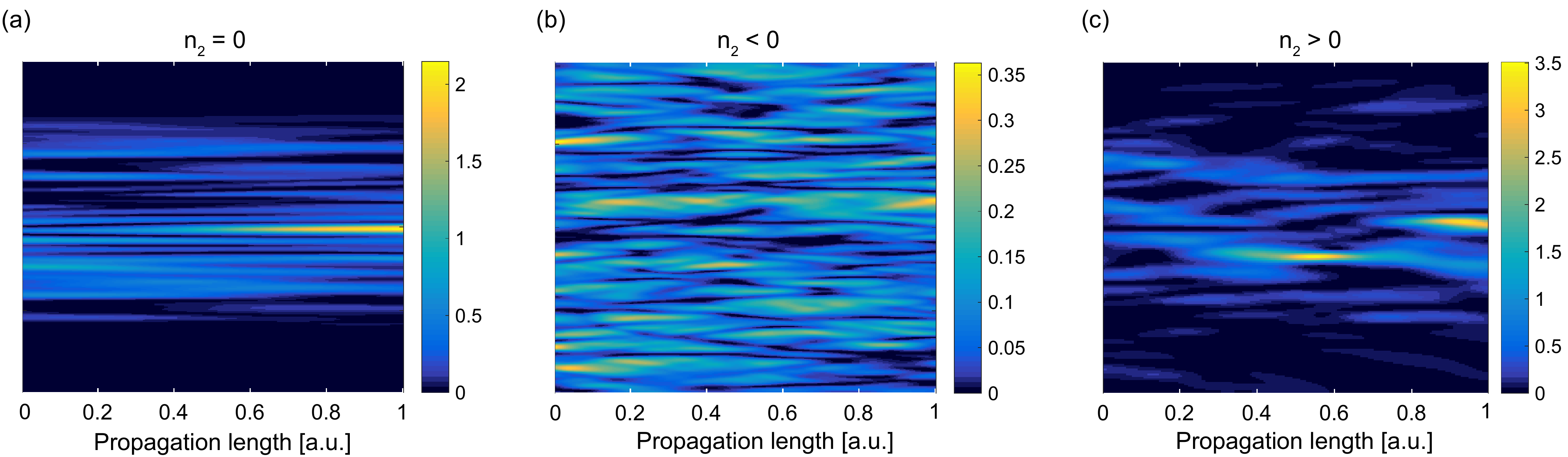}
\caption{Simulation results: 1D cuts through the propagation of 2D speckle fields, created by the same diffuser, inside the same (a) linear (b) defocusing nonlinear and (c) focusing nonlinear media as those in Fig.~3, but \textit{without} wavefront shaping.}
\label{UnoptSpecPropag}
\end{figure*}